\documentclass[cameraready]{Interspeech}
\usepackage{threeparttable}
\usepackage{enumitem}
\usepackage{booktabs} 
\usepackage{amssymb}
\usepackage{xcolor}
\usepackage{amsmath}

\title{VAD to the Bone: Ultra-Tiny \\Speech Activity Detection for Edge Deployment }

\author[affiliation={1,2}, equalcontribution]{Stephen}{Bauer}
\author[affiliation={1}, equalcontribution, correspondingauthor]{Sheila}{Seidel}
\author[affiliation={1}]{Shanza}{Iftikhar}
\author[affiliation={1}]{Scott}{Veidenheimer}
\author[affiliation={1}]{Gorkem}{Ulkar}

\address{
    $^1$ Analog Devices, Inc., USA \\
    $^2$ University of California, Los Angeles (UCLA), USA
}

\email{sheila.seidel@analog.com}


\usepackage{comment}
\usepackage[normalem]{ulem}

\begin{document}
\maketitle
\begin{center}
\small\itshape
Accepted for publication at INTERSPEECH 2026.
\end{center}


\begin{abstract}
Voice activity detection (VAD) triggers downstream speech processing in always-on systems under strict memory, latency, and compute constraints. Recent compact models report strong accuracy but rely on components that are not widely supported: learnable filterbanks, recurrent layers, or non-causal post-processing. We propose kiloVAD, designed for embedded inference using standard Mel features, CNN-only layers, and tunable context/spectral parameters. We introduce per-layer structured pruning with self-distillation and angle-based quantization-aware training (QAT) that outperforms standard QAT by 1-4\%. Evaluated per-frame under causal conditions, kiloVAD achieves 0.850 AUC on AVA-Speech with 2.1 k parameters and 200 ms context, establishing a new state of the art for causal, deployment-ready VAD.  
\end{abstract}

\section{Introduction}
Voice activity detection (VAD) is a critical front-end component for speech processing systems deployed on edge devices. Before any downstream processing such as speech recognition, speaker verification, or keyword spotting can engage, a VAD must first determine whether speech is present, enabling the device to keep power-hungry models dormant until needed. For always-on applications like smart speakers, hearables, and IoT devices, this decision must be made continuously with minimal latency and power consumption.  

Recent advances in deep learning have led to significant improvements in VAD robustness, particularly in noisy and real-world conditions~\cite{silero_vad, trvad, karan2024transformer, pyannote}. However, these models achieve their performance at the cost of millions of parameters, making them impractical for always-on, resource-constrained devices. To address this, recent work has produced compact VAD architectures. MarbleNet \cite{jia2021marblenetdeep1dtimechannel} introduced 1D time-channel separable convolutions at 91 k parameters. TinyVAD \cite{tinyvad} reduced this to 11.6 k parameters using a patchify module and CSPTiny layers. SincQDR \cite{sincqdr} achieved 8 k parameters with learnable sinc filters, while ResectNet \cite{resectNet} reached 4.5 k parameters by combining sinc convolutions with a GRU. SG-VAD \cite{sgvad} proposed stochastic gates for feature selection at 7.8 k parameters. Most recently, AtomicVAD \cite{atomicVAD} achieved 0.3 k parameters using a novel oscillatory activation function (GGCU).   

However, parameter count alone does not determine deployment feasibility. We identify four considerations for practical embedded use.   First, \textit{frontend compatibility}: several recent models, such as ResectNet, SincQDR, and AtomicVAD operate on raw audio with learnable filterbanks or internal spectrogram computation, requiring custom DSP implementations that cannot leverage hardware-accelerated Mel spectrogram extraction available on most embedded platforms. Moreover, Mel features are shared by downstream tasks such as keyword spotting and speech recognition, enabling a single frontend to serve multiple models in the audio pipeline. Second, \textit{architectural constraints}: some compact VADs employ specialized recurrent or gated units that introduce non-standard operations. For example, AtomicVAD's GGCU activation requires computing $\cos(\cdot)$ per activation, which is expensive on microcontrollers without hardware trigonometric computation support. ResectNet's GRU introduces dynamic control flow problematic for frameworks like TensorFlow Lite for Microcontrollers (TFLM). Third, \textit{detection latency:} prior work often optimizes for much higher latency. SG-VAD operates at the segment level (up to 100 seconds), incompatible with low-latency streaming. MarbleNet, TinyVAD, and AtomicVAD all report \SI{630}{\milli\second} input context, tripling latency compared to our \SI{200}{\milli\second} design. Fourth, \textit{causal evaluation}: as noted by \cite{atomicVAD}, ``reported values are not directly comparable because of two competing inference protocols'', yet most compact VADs, including AtomicVAD itself, report best results using non-causal sliding window inference with 87.5\% overlap. AtomicVAD's causal performance on AVA Speech \cite{ava_speech} achieves an Area Under the Curve (AUC) of 0.869, dropping notably from its non-causal result of 0.903, illustrating how evaluation protocol choice may overstate streaming performance.   

Beyond architecture, edge deployment demands aggressive model compression to fit within the memory and compute budgets of target hardware—budgets that vary widely across microcontroller classes. Two complementary techniques dominate this space: structured pruning, which removes entire channels or layers to reduce parameter counts, often combined with knowledge distillation to recover accuracy~\cite{jiang2023accurate}; and quantization-aware training (QAT), which enables inference with fixed-point arithmetic but typically requires careful handling to avoid degradation at low bit-widths~\cite{nguyen2020quantization}. 

In this work, we present kiloVAD, a deployment-oriented VAD designed to jointly address architecture, compression, and real-time constraints. Our main contributions include:   \begin{itemize}[nosep,leftmargin=*] 
\item A CNN-only architecture using standard Mel spectrogram features, designed for compatibility with embedded ML toolchains and aggressive structured pruning.  
\item A generalizable per-layer structured pruning optimization approach with self-distillation, enabling varying degrees of model-size reduction to suit application requirements.
\item A novel angle-aware self-distilling QAT approach that outperforms standard QAT by 1--4\% under INT4 quantization. 
\item Analysis of design tradeoffs including Mel bins, input context, and compression ratio versus performance.                 \end{itemize} 
We evaluate under strictly causal conditions with a \SI{200}{\milli\second} input context, achieving 0.850 AUC on AVA-Speech with 2.1 k parameters pre-quantization. To facilitate reproducibility, we provide pretrained weights and a web-based demo.\footnote{\url{https://huggingface.co/spaces/kiloVAD-demo/kiloVAD}}

\section{Deployment Oriented System Design}
\begin{figure}[t]
    \centering
    \includegraphics[width=0.92\linewidth]{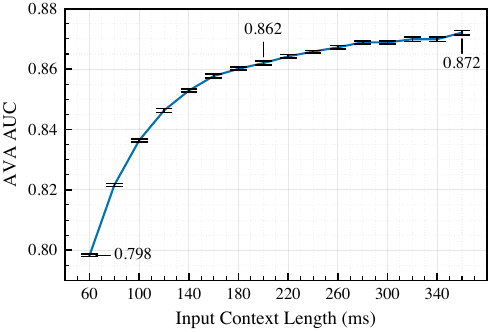}
    \caption{AVA-Speech AUC vs.\ input context length. Error bars show 95\% confidence intervals ($n$=10 seeds).   
  Performance plateaus near 200\,ms.}
    \label{fig:ctx_window}
\end{figure}
\subsection{Model Architecture}
The kiloVAD architecture is optimized for cross-platform embedded support under standard operator constraints. Our model processes Mel spectrogram features through depthwise separable convolutional blocks, global average pooling, and a linear classifier. We make four deployment-driven design choices: 

\textbf{Convolution-only design.} kiloVAD is designed for compatibility with embedded ML toolchains and supports export using only standard operations. We present a convolution-only backbone (all layers except the final classifier) operating on fixed Mel-spectrogram features and avoid non-standard components such as Sinc filters, GRUs, or novel activation functions that require custom kernels or introduce dynamic control flow. This design ensures static-graph compatibility and eliminates unsupported operators in lightweight embedded ML runtimes.

\textbf{Pruning-friendly adapter layer.} A $1 \times 1$ convolutional adapter projects mel features to an internal channel dimension. This decouples the fixed mel resolution from subsequent layer widths, enabling aggressive structured pruning of internal channels without modifying the input interface.  

\textbf{Global average pooling.} Rather than flattening features before classification (which ties parameter count to input length), we pool across the temporal dimension. This allows the same weights to operate on different input context lengths without architectural changes or additional parameters.   

\textbf{Amplitude-agnostic preprocessing.} We normalize each mel frequency bin to zero mean and unit variance across the input window before passing features to the model. This per-frame normalization forces the model to learn relative spectral patterns rather than absolute energy levels, improving robustness across diverse recording conditions and microphone gains. The normalization adds minimal overhead: computing mean and variance over 21 time steps for 64 mel bins.

The architecture comprises: an adapter layer ($n_{\text{mels}} \rightarrow 128$), a depthwise separable block (temporal kernel 11), two $1 \times 1$ projection layers (128$\rightarrow$64$\rightarrow$64 channels), a residual block (kernel 17), a dilated block (kernel 29, dilation 2), a pointwise conv, global average pooling, and a binary classifier.  

\subsection{Deployment-Driven Feature and Context Selection}
\textbf{Input context.} The input context length directly determines detection latency. Figure~\ref{fig:ctx_window} shows AUC across context lengths from 60\,ms to 360\,ms. Performance rises sharply from 0.798 at 60\,ms to 0.862 at 200\,ms, then plateaus, consistent with the 4--5\,Hz syllable rate of spontaneous speech~\cite{speechIntuition}. A 200\,ms input context captures roughly one syllable, providing sufficient acoustic evidence while also providing a significant latency reduction over the 630\,ms latency required by MarbleNet~\cite{jia2021marblenetdeep1dtimechannel}, TinyVAD~\cite{tinyvad}, and AtomicVAD~\cite{atomicVAD}.         

\begin{figure}[t]
    \centering
    \includegraphics[width=0.92\linewidth]{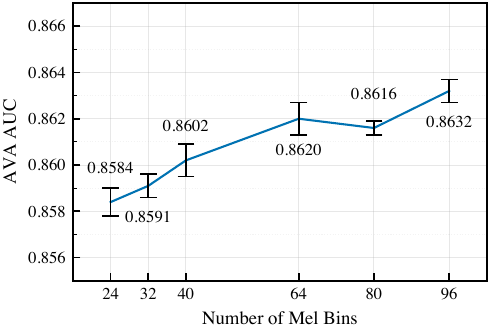}
    \caption{AVA-Speech AUC vs.\ Mel bin count. Error bars show 95\% confidence intervals ($n$=10 seeds).   
  Performance degrades gracefully with fewer bins.}
    \label{fig:mel_bins}
\end{figure}

\textbf{Mel resolution.} Embedded DSP libraries vary in supported Mel configurations.  For a 200\,ms input context, we evaluate Mel resolutions from 24 to 96 bins (Figure~\ref{fig:mel_bins}) and find performance degrades gracefully: reducing from 64 to 32 bins costs only 0.003 AUC points. We default to 64 bins but support lower resolutions for platforms with different DSP constraints. 

\section{Compression Methods}
\subsection{Per-Layer Structured Pruning Strategy}
We use structured pruning via \texttt{torch-pruning}~\cite{depgraphPruning}, which constructs a dependency graph to identify parameter groups that must be pruned together to preserve functional structure. We apply magnitude-based pruning with $\ell_2$-norm importance to systematically remove channels contributing least to the output.

Rather than applying a uniform global pruning ratio, we allow each layer to have an independent pruning ratio. We search over these per-layer ratios using Optuna~\cite{optuna} with a multi-objective formulation: simultaneously  minimizing (1) the False Positive Rate (FPR) at a True Positive Rate (TPR) of 0.95 on validation data, and (2) total parameter count. This produces a Pareto front of pruning configurations, allowing us to select the optimal model for any target size.                          

After pruning, we fine-tune for 8 epochs using self-distillation: the unpruned model serves as the teacher, and we  minimize a weighted combination of cross-entropy loss and KL divergence between student and teacher logits. This recovers accuracy lost during pruning without requiring a separately trained teacher.

\subsection{Self-Distilling Quantization-Aware Training}

Quantization complements pruning for embedded deployment, enabling faster inference, lower energy, and reduced memory via fixed-point arithmetic. As noted in \cite{anderson2017high, zhang2019tnt} angular quantization error dominates at low bit-widths. Prior work \cite{li2023} addresses this via knowledge distillation, learning quantized features to match full-precision teacher directions. We go further: we freeze the full-precision classifier and treat its class weight vectors as fixed prototypes. Our loss pulls penultimate feature vectors toward their corresponding class weight vectors while repelling them from non-target weight vectors, directly optimizing feature--weight geometry without a teacher. Unlike angular-margin softmax methods \cite{liu2017sphereface,wang2018cosface,deng2019arcface}, which modify full-precision training, we specifically target quantization-induced angle errors. Our loss operates on the quantized graph with soft-to-hard annealing inspired by \cite{agustsson2017softtohard} and regularization to stabilize angle drift.    
    
\textbf{Setup.} For a mini-batch $\{(x_i,y_i)\}_{i=1}^B$ with $y_i \in \{1,\ldots,C\}$, where $C$ is the number of classes,
let the backbone have weights $U$ and the classifier be a \emph{frozen}
full-precision matrix $W^{\mathrm{FP}} \in \mathbb{R}^{C \times d}$ with rows
$w_c^{\mathrm{FP}} \in \mathbb{R}^d$ (one per class).
We quantize \emph{only} the backbone: $\tilde U = Q_W(U)$.
For each sample, the penultimate feature is $f_i = f_{\tilde U}(x_i) \in \mathbb{R}^d$,
so $f_i$ and each $w_c^{\mathrm{FP}}$ live in the same $d$-dimensional space.
We use the cosine-similarity function and do not backpropagate through $W^{\mathrm{FP}}$.

\textbf{Objective.} The training loss is the mini-batch mean of an align–repel objective:
\begin{align}
\label{eq:loss-cossim}
\mathcal{L}
= \frac{1}{B}\sum_{i=1}^{B} \Big[
    &\underbrace{1 - \cos\big(f_i,\, w_{y_i}^{\mathrm{FP}}\big)}_{\text{align to target}} \notag \\
    &\quad +\; \lambda\,\underbrace{\phi\big(\{\cos(f_i, w_c^{\mathrm{FP}})\}_{c \in \{1,\ldots,C\}\setminus\{y_i\}}\big)}_{\text{repel from non-targets}}
\Big],
\end{align}
\vspace{-1 em}

with a hinge-style repulsion
\vspace{-0.5em}

\begin{equation} 
\label{eq:repel-cossim}
\phi\big(\{s_c\}_{c\neq y_i}\big) =\max\bigl\{0,\, \max_{c\neq y_i} s_c \bigr\}
\end{equation}
The first term in Eq.~(1) aligns the feature vector with the target class weight vector; the second term in Eq.~(1) uses the hinge penalty defined in Eq.~(2), which operates on cosine similarities $s_c = \cos(f_i, w_c^{\mathrm{FP}})$ between $f_i$ and all non-target prototypes $w_c^{\mathrm{FP}}$ with $c \neq y_i$. The $\max_{c \neq y_i} s_c$ term selects the non-target class whose prototype is most aligned with $f_i$, and the outer $\max\{0, \cdot\}$ clamps this value at zero, so the penalty is non-zero only when a non-target class is aligned with $f_i$. Thus the second term in Eq.~(1) penalizes features that are too close in angle to incorrect prototypes and encourages an angular margin between the target and all other classes. In our VAD setting $C=2$ so this term penalizes similarity between $f_i$ and the only other class prototype. For aggressive quantization, we use INT4 backbone weights and show that our angle-based self-distilling QAT substantially outperforms standard STE-based QAT~\cite{hubara2017qnn}
\section{Experimental Setup}
\textbf{Training data.} We train on LibriSpeech train-clean-100~\cite{librispeech} with three noise conditions: 25\% clean speech, 25\% mixed with synthetic wind noise~\cite{windnoise_sim} at $-5$\,dB SNR, and 50\% mixed with DNS Challenge noise~\cite{dns_challenge} at SNRs of $\{-10, -5, 0, 5, 10\}$\,dB, with half of these samples including simulated room reverberation. This mixture exposes the model to both environmental noise (wind) and diverse acoustic interference (DNS). For LibriSpeech, tightly aligned annotations were obtained using the Montreal Forced Aligner  \cite{montrealaligner} as provided by \cite{libri_annotations}.  
\begin{figure}[t]
    \centering
    \includegraphics[width=0.92\linewidth]{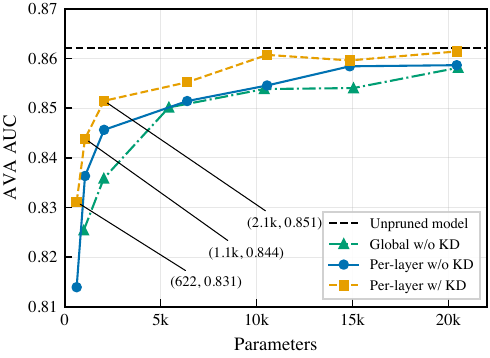}
    \caption{AVA-Speech AUC vs.\ parameter count for pruned models. Per-layer pruning outperforms uniform global pruning, especially at low parameter counts. Knowledge distillation (KD)   
  improves performance at all compression levels. Labeled points: 622 parameters retains 0.831 AUC; 2.1 k parameters achieves 0.851 AUC, within 1.3\% of baseline.}
    \label{fig:pruning}
\end{figure}

\textbf{Training protocol.} We use SGD with momentum 0.9, Nesterov acceleration, and weight decay $8.75 \times 10^{-4}$. The learning rate follows a cyclic schedule: 4-epoch warmup to $3.5 \times 10^{-3}$, 16-epoch hold, then 20-epoch cosine decay to $10^{-5}$. We apply label smoothing ($\epsilon = 0.09$) and train for 40 epochs with batch size 512.

\textbf{Evaluation.} We evaluate on AVA-Speech~\cite{ava_speech}, which contains 15-minute YouTube clips with annotated speech segments across diverse acoustic conditions. Evaluating on a different domain than training tests generalization to unseen acoustic environments. Unlike prior work that uses non-causal sliding-window inference with 87.5\% overlap~\cite{tinyvad, sincqdr}, we evaluate under strictly causal conditions: each 200\,ms frame is classified independently using that audio, with no future context or temporal smoothing. We report frame-level AUC and best F1 score at the optimal threshold, with  95\% confidence intervals computed over 10 independent training runs with different random seeds.   

\section{Results}

\begin{table*}[ht]                                                                                                                                           
  \centering                                                                                                                                                   
  \small                                                                                                                                                       
  \setlength{\tabcolsep}{2.5pt}                                                                                                                                
  \begin{threeparttable}                                                                                                                                       
  \caption{Comparison of kiloVAD to prior art. kiloVAD is the only model satisfying all four deployment requirements (R1--R4). For kiloVAD, 95\% CIs shown     
  (full: $n$=10; pruned: $n$=8, as 2 seeds exhibited layer collapse at extreme compression).}                                                                  
  \label{tab:main_results}                                                                                                                                     
  \begin{tabular}{lcccccccc}                                                                                                                                   
  \toprule                                                                                                                                                     
  \textbf{Model} & \textbf{R1: Frontend} & \textbf{R2: Portable} & \textbf{R3: Low} & \textbf{R4: Causal} & \textbf{Params} & \textbf{Input Ctx.} & \textbf{Total Ctx.} & \textbf{AUC} \\                                                                                                                                      
   & \textbf{Compat.} & \textbf{Ops} & \textbf{Latency} & \textbf{Eval} & (K) & (ms) & (ms) & (AVA) \\                                                         
  \midrule                                                                                                                                                     
  kiloVAD (full) & \checkmark & \checkmark & \checkmark & \checkmark & 81.1 & 200 & 200 & 0.862 $\pm$ 0.001 \\                                                 
  kiloVAD (pruned) & \checkmark & \checkmark & \checkmark & \checkmark & 2.1 & 200 & 200 & 0.850 $\pm$ 0.007 \\                                                
  \midrule                                                                                                                                                     
  MarbleNet~\cite{jia2021marblenetdeep1dtimechannel} & \checkmark & \checkmark & -- & \checkmark & 91 & 630 & 630 & 0.850 \\                                   
  TinyVAD~\cite{tinyvad} & \checkmark & \checkmark & -- & --\tnote{a} & 11.6 & 630 & 1181\tnote{a} & 0.864 \\                                                  
  SincQDR~\cite{sincqdr} & --\tnote{b} & --\tnote{c} & -- & --\tnote{a} & 8 & 25 & 1181\tnote{a} & 0.914 \\                                                    
  ResectNet~\cite{resectNet} & --\tnote{b} & --\tnote{d} & \checkmark & \checkmark & 4.5 & 40 & 200 & 0.886 \\                                                 
  AtomicVAD~\cite{atomicVAD} & --\tnote{b} & --\tnote{e} & -- & \checkmark & 0.3 & 630 & 630 & 0.869 \\                                                        
  \bottomrule                                                                                                                                                  
  \end{tabular}                                                                                                                                                
  \vspace{0.5em}                                                                                                                                               
  \noindent\footnotesize                                                                                                                                       
  \begin{tabular}{@{}p{0.48\linewidth}@{\hspace{0.04\linewidth}}p{0.48\linewidth}@{}}                                                                          
  \textsuperscript{a}~Uses non-causal sliding-window inference with 87.5\% overlap. &                                                                          
  \textsuperscript{d}~Uses GRU for temporal modeling. \\                                                                                                       
  \textsuperscript{b}~Operates on raw audio (not Mel-compatible). &                                                                                            
  \textsuperscript{e}~Uses GGCU activation (cos() per activation). \\                                                                                          
  \textsuperscript{c}~Uses learnable sinc bandpass filters. & \\                                                                                               
  \end{tabular}                                                                                                                                                
  \end{threeparttable}                                                                                                                                         
  \end{table*}                                
\textbf{Structured pruning.} Figure~\ref{fig:pruning} compares pruning strategies across compression levels. Per-layer pruning ratios, optimized via multi-objective search, consistently outperform uniform global pruning, particularly below 5 k parameters where global pruning degrades sharply. Notably, global pruning fails entirely below ~2 k parameters (the curve terminates) due to layer collapse, while per-layer pruning remains stable down to 622 parameters. Knowledge distillation from the unpruned teacher provides AUC gains at all compression levels. At 2.1 k parameters, per-layer pruning with KD achieves 0.851 AUC, within 1.3\% of the unpruned baseline while reducing parameters by 97.4\% and MACs from 1.7 M to 44 k per inference. We feature the 2.1 k configuration in Table~\ref{tab:main_results}.  Pruning ratios were optimized on a single seed (the model studied in Figure~\ref{fig:pruning}) and transferred to 10 independent models; 2 failed under this aggressive compression due to layer collapse, yielding $n$=8 for confidence intervals in Table~\ref{tab:main_results}. This transfer protocol tests generalization of the pruning configuration rather than per-model tuning.

\textbf{Comparison to prior work.} Table~\ref{tab:main_results} compares kiloVAD against recent compact VAD models. Critically, reported AUC values are not  directly comparable across methods: SincQDR and TinyVAD employ non-causal sliding-window inference with 87.5\% overlap, which substantially inflates performance. AtomicVAD reports 0.903 AUC with this protocol but only 0.869 under causal evaluation~\cite{atomicVAD}. Under strictly causal conditions with a 200\,ms input context, 3$\times$ shorter than MarbleNet, TinyVAD, and AtomicVAD, our pruned model (F1: $0.783 \pm 0.004$) matches MarbleNet (0.850 AUC) at 43$\times$ fewer parameters while maintaining TFLM portability. Our full model achieves 0.862 AUC (F1: $0.796 \pm 0.001$). As shown in Figure~\ref{fig:ctx_window}, extending input context to 360\,ms yields 0.872 AUC, surpassing both AtomicVAD's causal result (0.869) and TinyVAD's non-causal result (0.864), despite using nearly half the context (360\,ms vs.\ 630\,ms) and only standard CNN operations.      

\begin{figure}[t]
    \centering
    \includegraphics[width=0.9\linewidth]{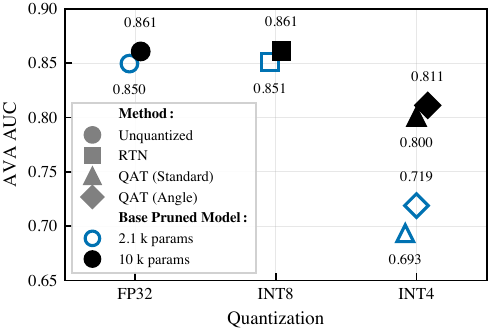}
    \caption{ AVA-Speech AUC for FP32, INT8, and INT4 models. Colors denote quantization method; open vs. filled markers indicate 2.1 k vs. 10 k parameters.}
    \label{fig:qat}
\end{figure}

\textbf{Quantization.} We evaluate our angle-aware self-distillation method on two pruned variants: 10 k and 2.1 k parameters (Figure~\ref{fig:qat}). Post-training INT8 quantization with round-to-nearest (RTN) is essentially lossless: the 10 k model maintains 0.861 AUC and the 2.1 k model remains at 0.851 AUC, matching their FP32 baselines. This confirms that 8-bit weights and activations can be deployed on embedded targets without sacrificing accuracy.   

For more aggressive INT4 quantization, our angle-aware self-distilling QAT outperforms standard straight-through estimator (STE) QAT~\cite{hubara2017qnn}. Standard INT4 QAT yields 0.800 AUC for the 10 k model and 0.693 AUC for the 2.1 k model. Angle-aware QAT improves these to 0.811 and 0.719 AUC, respectively, a 1--4\% relative improvement depending on model size. This demonstrates that explicit optimization of quantization-induced angular error provides meaningful gains under aggressive compression, without requiring a separate teacher model.

\section{Discussion}

We presented kiloVAD, a VAD co-designed from the ground up for aggressive compression and microcontroller deployment. Our architecture, featuring a 1$\times$1 adapter layer to decouple input resolution from internal channels, global average pooling for context-flexible inference, and a fully convolutional backbone, was specifically engineered to support extreme structured pruning while remaining TFLM-compatible. Per-layer pruning with multi-objective optimization reduces the model from 81 k to 2.1 k parameters (97.4\% reduction) and from 1.7 M to 44 k MACs, while retaining 0.850 AUC, within 1.3\% of the unpruned baseline. This matches MarbleNet at 43$\times$ fewer parameters with 3$\times$ lower latency (200\,ms vs.\ 630\,ms). At 360\,ms context, still nearly half that of AtomicVAD (630\,ms), kiloVAD reaches 0.872 AUC, exceeding AtomicVAD's 0.869 with lower latency and only standard CNN operations. INT8 post-training quantization is lossless; for INT4, our angle-aware self-distilling QAT outperforms standard QAT by 1--4\%, demonstrating that explicit optimization of quantization-induced angular error is critical at extreme bit-widths. Unlike prior work that achieves small footprints through recurrent architectures, learnable filterbanks, or non-causal post-processing, kiloVAD is CNN-only, fixed-point compatible, stateless, and low-latency, addressing a gap in deployment-ready VAD for resource-constrained edge devices.
\section{Acknowledgments}
\ifcameraready
The authors thank Bibek Gupta and Rupesh Kumar for their help configuring on-premises machines and Azure training workflows, Adam Rowell for his help creating the online demo, and Kao Kitichotkul, Alessandro Ragano, and Henning Hasemann for their feedback on the manuscript. 
\else
\fi

\section{Generative AI Use Disclosure}
Claude (Anthropic) was used as a writing assistant for manuscript editing and as a programming aid during code development. The authors reviewed and validated all outputs and retain full responsibility for the work presented.

\bibliographystyle{IEEEtran}
\bibliography{mybib}

@misc{jia2021marblenetdeep1dtimechannel,
      title={MarbleNet: Deep 1D Time-Channel Separable Convolutional Neural Network for Voice Activity Detection}, 
      author={Fei Jia and Somshubra Majumdar and Boris Ginsburg},
      year={2021},
      eprint={2010.13886},
      archivePrefix={arXiv},
      primaryClass={eess.AS},
      url={https://arxiv.org/abs/2010.13886}, 
}

@inproceedings{resectNet,
  author    = {Köpüklü, Okan and Taseska, Maja},
  title     = {ResectNet: An Efficient Architecture for Voice Activity Detection on Mobile Devices},
  booktitle = {Proceedings of Interspeech 2022},
  year      = {2022},
  month     = {09},
  pages     = {5363--5367},
  doi       = {10.21437/Interspeech.2022-820}
}

@article{hubara2017qnn,
  title   = {Quantized Neural Networks: Training Neural Networks with Low Precision Weights and Activations},
  author  = {Hubara, Itay and Courbariaux, Matthieu and Soudry, Daniel and El-Yaniv, Ran and Bengio, Yoshua},
  journal = {Journal of Machine Learning Research},
  volume  = {18},
  number  = {187},
  pages   = {1--30},
  year    = {2018}
}

@INPROCEEDINGS{tinyvad,
  author={Chae, Hwabyeong and Lee, Sunggu},
  booktitle={ICASSP 2024 - 2024 IEEE International Conference on Acoustics, Speech and Signal Processing (ICASSP)}, 
  title={Small-Footprint Convolutional Neural Network with Reduced Feature Map for Voice Activity Detection}, 
  year={2024},
  volume={},
  number={},
  pages={12266-12270},
  doi={10.1109/ICASSP48485.2024.10446312}}

@inproceedings{dns_challenge,                               title={The INTERSPEECH 2020 Deep Noise Suppression Challenge: Datasets, Subjective Testing Framework, and      Challenge Results},                                         author={Reddy, Chandan KA and others},                      booktitle={Interspeech},                                    year={2020}                                                 }

@inproceedings{optuna, 
title={Optuna: A Next-generation Hyperparameter Optimization Framework}, 
author={Akiba, Takuya and Sano, Shotaro and Yanase, Toshihiko and Ohta, Takeru and Koyama, Masanori},
booktitle={KDD},
year={2019}
}

@article{speechIntuition,
  title={Temporal properties of spontaneous speech—a syllable-centric perspective}, 
  author={Greenberg, Steven and Carvey, Hannah and Hitchcock, Leah and Chang, Shuangyu}, 
  journal={Journal of Phonetics},
  volume={31},                                        number={3-4},                                       pages={465--485},                                   year={2003},                                        publisher={Elsevier}                                }

@INPROCEEDINGS{trvad,
author={Zhao, Yifei and Champagne, Benoit},
booktitle={2022 IEEE 32nd International Workshop on Machine Learning for Signal Processing (MLSP)}, 
title={An Efficient Transformer-Based Model for Voice Activity Detection}, 
year={2022},
volume={},
number={},
pages={1-6},
doi={10.1109/MLSP55214.2022.9943501}}

@inproceedings{sgvad,
  title={SG-VAD: Stochastic Gates Based Speech Activity Detection},
  author={Svirsky, Jonathan and Lindenbaum, Ofir},
  booktitle={ICASSP 2023-2023 IEEE International Conference on Acoustics, Speech and Signal Processing (ICASSP)},
  pages={1--5},
  year={2023},
  organization={IEEE}
}

@inproceedings{depgraphPruning,
  title={Depgraph: Towards any structural pruning},
  author={Fang, Gongfan and Ma, Xinyin and Song, Mingli and Mi, Michael Bi and Wang, Xinchao},
  booktitle={Proceedings of the IEEE/CVF Conference on Computer Vision and Pattern Recognition},
  pages={16091--16101},
  year={2023}
}

@INPROCEEDINGS{windnoise_sim,
  author={Mirabilii, Daniele and Lodermeyer, Alexander and Czwielong, Felix and Becker, Stefan and Habets, Emanuël A.P.},
  booktitle={2022 International Workshop on Acoustic Signal Enhancement (IWAENC)}, 
  title={Simulating Wind Noise with Airflow Speed-Dependent Characteristics}, 
  year={2022},
  volume={},
  number={},
  pages={1-5},
  doi={10.1109/IWAENC53105.2022.9914785}}

@inproceedings{jiang2023accurate,
title     = {Accurate and Structured Pruning for Efficient Automatic Speech Recognition}, 
author    = {Jiang, Huiqiang and Zhang, Li Lyna and Li, Yuang and Wu, Yu and Cao, Shijie and Cao, Ting and Yang,  
  Yuqing and Li, Jinyu and Yang, Mao and Qiu, Lili}, 
  booktitle = {Proc. Interspeech},  
  pages     = {4104--4108}, 
  year      = {2023}, 
  doi       = {10.21437/Interspeech.2023-809}                 }

@inproceedings{nguyen2020quantization, 
  title     = {Quantization Aware Training with Absolute-Cosine Regularization for Automatic Speech Recognition}, author    = {Nguyen, Hieu Duy and Alexandridis, Anastasios and Mouchtaris, Athanasios}, 
  booktitle = {Proc. Interspeech},
  pages     = {3366--3370}, 
  year      = {2020}, 
  doi       = {10.21437/Interspeech.2020-1991}  
  }

@inproceedings{ava_speech,
  title     = {{AVA-Speech: A Densely Labeled Dataset of Speech Activity in Movies}},
  author    = {Chaudhuri et al., Sourish},
  booktitle = {Proceedings of Interspeech},
  year      = {2018},
  pages     = {1239--1243},
  doi       = {10.21437/Interspeech.2018-2028},
  url       = {https://www.isca-speech.org/archive/Interspeech_2018/abstracts/2028.html}
}

@inproceedings{montrealaligner,
  author={McAuliffe, Michael and Socolof, Michaela and Mihuc, Sarah and Wagner, Michael and Sonderegger, Morgan},
  title={{Montreal Forced Aligner: Trainable Text-Speech Alignment Using Kaldi}},
  year=2017,
  booktitle={Proc. Interspeech 2017},
  pages={498--502},
  doi={10.21437/Interspeech.2017-1386}
}

@inproceedings{librispeech,
  title={Librispeech: an asr corpus based on public domain audio books},
  author={Panayotov, Vassil and Chen, Guoguo and Povey, Daniel and Khudanpur, Sanjeev},
  booktitle={2015 IEEE international conference on acoustics, speech and signal processing (ICASSP)},
  pages={5206--5210},
  year={2015},
  organization={IEEE}
}

@INPROCEEDINGS{pyannote,
  author={Bredin et al., Hervé},
  booktitle={ICASSP 2020 - 2020 IEEE International Conference on Acoustics, Speech and Signal Processing (ICASSP)}, 
  title={Pyannote.Audio: Neural Building Blocks for Speaker Diarization}, 
  year={2020},
  volume={},
  number={},
  pages={7124-7128},
  doi={10.1109/ICASSP40776.2020.9052974}}

@misc{libri_annotations,
      title={Speech Model Pre-training for End-to-End Spoken Language Understanding}, 
      author={Loren Lugosch and Mirco Ravanelli and Patrick Ignoto and Vikrant Singh Tomar and Yoshua Bengio},
      year={2019},
      eprint={1904.03670},
      archivePrefix={arXiv},
      primaryClass={eess.AS},
      url={https://arxiv.org/abs/1904.03670}, 
}

@article{anderson2017high,
  title   = {The High-Dimensional Geometry of Binary Neural Networks},
  author  = {Anderson, Alexander G. and Berg, Cory P.},
  journal = {arXiv preprint arXiv:1705.07199},
  year    = {2017}
}

@misc{silero_vad,
  title        = {Silero VAD: Pre-trained enterprise-grade Voice Activity Detector (VAD) models},
  author       = {Silero Team},
  year         = {2025},
  howpublished = {\url{https://github.com/snakers4/silero-vad}},
  note         = {Version 6.0.0, released August 26, 2025}
}

@inproceedings{zhang2019tnt,
  author    = {Zhang, T. and Zhu, L. and Zhao, Q. and Shin, K.},
  title     = {Neural Networks Weights Quantization: Target None-retraining Ternary (TNT)},
  booktitle = {2019 Fifth Workshop on Energy Efficient Machine Learning and Cognitive Computing -- NeurIPS Edition (EMC2-NIPS)},
  pages     = {62--65},
  year      = {2019},
  month     = dec,
  publisher = {IEEE}
}

@article{sincqdr,
  title={SincQDR-VAD: A Noise-Robust Voice Activity Detection Framework Leveraging Learnable Filters and Ranking-Aware Optimization},
  author={Wang et al., Chien-Chun},
  journal={arXiv preprint arXiv:2508.20885},
  year={2025}
}

@inproceedings{karan2024transformer,
  title={A transformer-based voice activity detector},
  author={Karan et al., Biswajit},
  booktitle={Interspeech},
  volume={2024},
  pages={3819--3823},
  year={2024}
}

@ARTICLE{li2023,
  author={Li, Zhijian and Yang, Biao and Yin, Penghang and Qi, Yingyong and Xin, Jack},
  journal={IEEE Access}, 
  title={Feature Affinity Assisted Knowledge Distillation and Quantization of Deep Neural Networks on Label-Free Data}, 
  year={2023},
  volume={11},
  number={},
  pages={78042-78051},
  doi={10.1109/ACCESS.2023.3297890}}

@inproceedings{agustsson2017softtohard,
  title={Soft-to-Hard Vector Quantization for End-to-End Learning Compressible Representations},
  author={Agustsson, Eirikur and Mentzer, Fabian and Tschannen, Michael and Cavigelli, Lukas and Timofte, Radu and Benini, Luca and Van Gool, Luc},
  booktitle={NeurIPS},
  pages={1141--1151},
  year={2017}
}

@inproceedings{deng2019arcface,
  title={ArcFace: Additive Angular Margin Loss for Deep Face Recognition},
  author={Deng, Jiankang and Guo, Jia and Xue, Niannan and Kotsia, Irene and Zafeiriou, Stefanos},
  booktitle={CVPR},
  year={2019}
}

@inproceedings{liu2017sphereface,
  title={SphereFace: Deep Hypersphere Embedding for Face Recognition},
  author={Liu, Weiyang and Wen, Yandong and Yu, Zhiding and Li, Ming and Raj, Bhiksha and Song, Le},
  booktitle={CVPR},
  pages={6738--6746},
  year={2017}
}

@inproceedings{wang2018cosface,
  title={CosFace: Large Margin Cosine Loss for Deep Face Recognition},
  author={Wang, Hao and Wang, Yitong and Zhou, Zheng and Ji, Xing and Gong, Dihong and Zhou, Jingchao and Li, Zhifeng and Liu, Wei},
  booktitle={CVPR},
  pages={5265--5274},
  year={2018}
}

@article{atomicVAD,
  title={AtomicVAD: A Tiny Voice Activity Detection Model for Efficient Inference in Intelligent IoT Systems},
  author={Soto-Vergel, Angelo J and Sankaran, Prashant and Velez, Juan C and Amaya-Mier, Rene and Ramirez-Rios, Diana},
  journal={Internet of Things},
  pages={101822},
  year={2025},
  publisher={Elsevier}
}

\end{document}